\documentclass[conference,a4paper]{APSIPA2020}
\usepackage{multirow}
\usepackage[dvips]{graphicx}
\usepackage{amsmath}
\usepackage[psamsfonts]{amssymb}
\usepackage{amsxtra}
\usepackage{threeparttable}

\begin{document}

\title{End-to-end Music-mixed Speech Recognition}

\author{%
\authorblockN{%
Jeongwoo Woo\authorrefmark{1}, 
Masato Mimura\authorrefmark{1}, 
Kazuyoshi Yoshii\authorrefmark{1} and 
Tatsuya Kawahara\authorrefmark{1}
}
\authorblockA{%
\authorrefmark{1}
Kyoto University, Kyoto, Japan \\
\{woo, mimura, yoshii, kawahara\}@sap.ist.i.kyoto-u.ac.jp
}
}

\maketitle
\thispagestyle{empty}

\begin{abstract}
Automatic speech recognition (ASR) in multimedia content is one of the promising applications, but speech data in  this kind of content are frequently mixed with background music, which is harmful for the performance of ASR.
In this study, we propose a method for improving ASR with background music based on time-domain source separation. 
We utilize Conv-TasNet as a separation network, which has achieved state-of-the-art performance for multi-speaker source separation, to extract the speech signal from a speech-music mixture in the waveform domain. 
We also propose joint fine-tuning of a pre-trained Conv-TasNet front-end with an attention-based ASR back-end using both separation and ASR objectives.
We evaluated our method through ASR experiments using speech data mixed with background music from a wide variety of Japanese animations.
We show that time-domain speech-music separation drastically improves ASR performance of the back-end model trained with mixture data, and the joint optimization yielded a further significant WER reduction.
The time-domain separation method outperformed a frequency-domain separation method, which reuses the phase information of the input mixture signal, both in simple cascading and joint training settings.
We also demonstrate that our method works robustly for music interference from classical, jazz and popular genres.\end{abstract}

\vspace{-4pt}
\section{Introduction}
\vspace{-4pt}
Automatic speech recognition (ASR) for multimedia content such as movies, dramas, broadcast news and other online videos is useful for automatic subtitle generation.
Speech signals in these kinds of real-world multimedia content are frequently mixed with background music in order to make listeners immersed in.
However, the music in speech signal causes significant performance degradation in ASR \cite{Vanroose2003}\cite{Hughes2012}.
\par
Despite its importance, there is limited research on music-mixed ASR compared to noisy and reverberant ASR.
In a small number of existing approaches, it was tackled by removing music interference with unsupervised separation methods such as NMF \cite{RobustNMF} and robust PCA \cite{RobustPCA} or denoising autoencoders \cite{CDAE}\cite{DAE}.
In \cite{RobustNMF}, NMF was employed for extracting speech signal from a speech-music mixture and ASR was performed with a GMM-HMM acoustic model trained on clean data.
Zhao et al. \cite{CDAE} trained a denoising autoencoder using mixture data as input and clean speech as target.
They used a hybrid DNN-HMM acoustic model for ASR. In general, music-mixed ASR is very challenging, since the music interference is highly non-stationary and at low signal-to-noise ratios.
\par
Recently a fully-convolutional neural network called a convolutional time-domain audio separation network (Conv-TasNet) \cite{ConvTasNet}, which operates in the waveform domain, has shown to achieve an excellent performance for multi-speaker source separation.
Due to its high performance and flexibility, it was also adopted for many other tasks.
Kinoshita et al. \cite{NoiseReduction} used a Conv-TasNet for speech denoising and dereverberation and achieved better ASR performance than a frequency-domain counterpart method.
D\'efossez et al. \cite{SingingVoice} used a Conv-TasNet and its variant for singing voice separation.
\par
In this paper, motivated by the recent progress in the time-domain source separation mentioned above, we investigate the use of a Conv-TasNet for the purpose of improving speech recognition with background music interference.
Thus, we adopt a Conv-TasNet for the speech-music separation task and utilize an attention-based  \cite{AttentionASR} ASR model for the speech recognition task, whereas the previous studies utilized an HMM-based ASR model.
\par
This allows us to combine the Conv-TasNet front-end with the attention-based ASR back-end to form an end-to-end music-mixed ASR system that directly operates on raw waveform features. 
 A similar network for the multi-speaker source separation task has been used in \cite{JointConvTasNet}. 
 We retrain the pre-trained front-end and the back-end models jointly and evaluate our method through ASR experiments using speech data mixed with background music from a wide variety of Japanese animations.
\par
The rest of the paper is organized as follows. We introduce speech-music separation with Conv-TasNet in Section \MakeUppercase{\romannumeral 2}, describe joint training of Conv-TasNet and ASR model in Section \MakeUppercase{\romannumeral 3}, describe the experimental details in Section \MakeUppercase{\romannumeral 4}, show the results of the experiments and discussions in Section \MakeUppercase{\romannumeral 5}, and conclude the paper in Section \MakeUppercase{\romannumeral 6}.

\begin{figure*}[t]
    \centering
    \includegraphics[width=\textwidth]{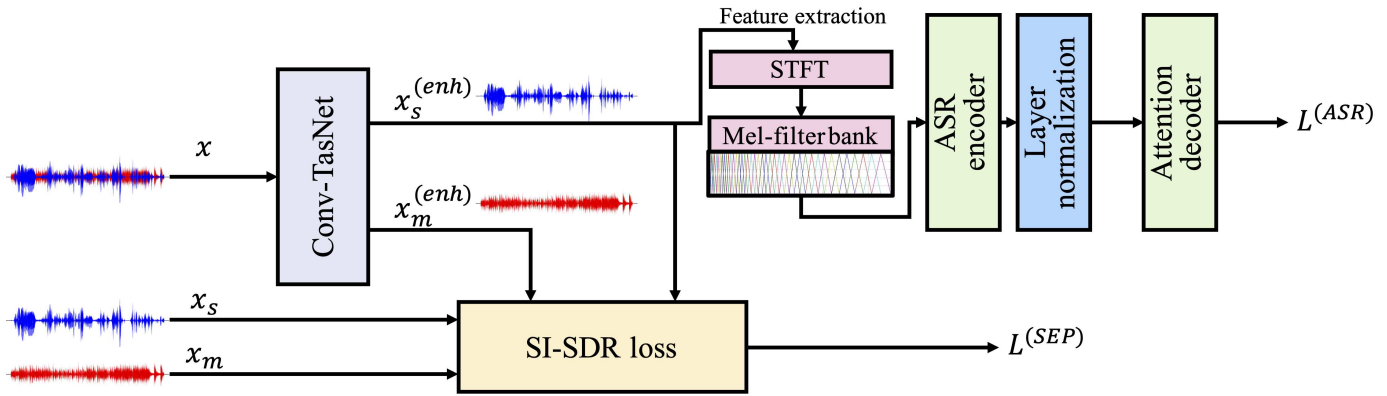}
    \caption{The proposed network architecture}
    \vspace{-18pt}
    \label{fig:model}
\end{figure*}

\vspace{-4pt}
\section{Speech-Music Separation and ASR}
\vspace{-4pt}
\subsection{Time-domain speech-music separation with Conv-TasNet}
\vspace{-4pt}
Conv-TasNet \cite{ConvTasNet} is a single-channel source separation model which generates waveforms for a fixed number of speakers from a mixture waveform. 
For speech-music separation, it outputs an estimated speech audio stream and an estimated music audio stream in the time domain, given an input mixture signal $\mathbf{x}$: 
\vspace{-4pt}
$$[\mathbf{x}_{s}^{(enh)}, \mathbf{x}_{m}^{(enh)}] = ConvTasNet(\mathbf{x}) \vspace{-5pt}$$

The model works directly on raw waveforms using an encoder and a decoder that can be learned in place of STFT and ISTFT. This makes it possible to obtain the phase information to reconstruct a waveform and to propagate the gradients from feature extraction to the waveform reconstruction part.
In the training stage, the scale-invariant-source-to-distortion ratio (SI-SDR) \cite{SI-SDR} loss is optimized directly on the time domain.\\
We refer to the loss for the Conv-TasNet as $L^{(SEP)}$,
\vspace{-4pt}
$$L^{(SEP)} = - \frac{\text{SI-SDR}_{s} + \text{SI-SDR}_{m}} {2} \vspace{-5pt}$$
where $\text{SI-SDR}_{s}$ and $\text{SI-SDR}_{m}$ are the SI-SDR loss of estimated speech and music, respectively.
\par
Note that Conv-TasNet was originally proposed for multi-speaker source separation, and usually requires permutation invariant training (PIT) \cite{uPIT} to solve permutation ambiguity of the output speech sources. Since the front-end outputs are speech and music in this work, we can fix the output order of the network and do not need permutation invariant training.
\vspace{-4pt}
\subsection{Attention-based ASR model}
\vspace{-4pt}
For ASR, we use an attention-based model.
We adopt an encoder-decoder architecture similar to \cite{AttentionASR}.
This architecture consists of two distinct networks of encoder and decoder.
The encoder transforms a sequence of acoustic features to a sequential representation, and the decoder predicts a label sequence from the encoded sequential representation using the attention mechanism.
We refer to the loss for the ASR component as $L^{(ASR)}$ which is the cross-entropy loss function with label smoothing.

\vspace{-4pt}
\section{Joint training of Conv-TasNet and ASR}
\vspace{-4pt}
We propose to combine the Conv-TasNet speech-music separation front-end with the attention-based ASR back-end, as shown in Fig. \ref{fig:model}. The input mixture $\mathbf{x}$ is separated by the front-end and the output speech audio stream is processed by the ASR back-end.
\par
Though music-mixed ASR can be performed by simply cascading the independently pre-trained front-end and back-end models, the speech-music separator produces unknown artifacts that degrade the performance of ASR.
According to \cite{JointConvTasNet}, such mismatches can be mitigated by joint fine-tuning the entire model.
\par
To propagate gradients from the ASR back-end to the front-end, we extract acoustic features for ASR directly from the speech waveform estimated by the front-end.
Specifically, log Mel-scale filterbank (lmfb) features obtained by applying an lmfb to an amplitude spectrogram generated from the speech waveform using STFT are used as acoustic features for the ASR model.
Since these processes are all differentiable, gradients can be propagated from the ASR back-end to the front-end.
In this end-to-end model, however, the acoustic features for ASR cannot be normalized beforehand. 
Thus, we insert the layer normalization \cite{LayerNorm} behind the encoder of the ASR back-end.
The losses for the front-end and the back-end are combined as
\vspace{-4pt}
$$ L = L^{(SEP)} + \alpha L^{(ASR)} \vspace{-5pt}$$
where $\alpha$ is an empirically chosen weight for the ASR loss.

\begin{table*}[t]
\small
\centering
\caption{WER and SDR on CSJ-anime for different variants of fine-tuning of joint model.}

\label{tb1}

\begin{tabular}{c|cc|c|cccc|ccc}
\multicolumn{11}{r}{WER: Word error rate, SDR: Signal-to-distortion ratio}\\
\hline \hline
& \multicolumn{2}{c|}{fine-tune} & \multicolumn{5}{c|}{WER(\%)} & \multicolumn{3}{c}{SDR(dB)} \\
Model & SEP & ASR & Clean & 5 dB & 0 dB & -5 dB & avg & 5 dB & 0 dB & -5 dB\\
\hline \hline
Clean & -- & -- & 11.25  & 46.62 & 78.96 & 93.33 & 72.97 & -- & -- & --\\
Mixture & -- & -- & 12.29 & 15.26 & 19.63 & 31.57 & 22.15 & -- & -- & --\\
\hline \hline
Frequency Domain	& -- & -- & 11.24  & 16.03 & 22.76 & 37.84 & 25.54 & 11.95 & 9.69 & 7.23\\
separation Model	& \checkmark & -- & 11.73 & 14.91 & 18.85 & 30.15 & 21.30 & 11.07 & 8.76 & 6.21\\
+				& -- & \checkmark & 11.70 & 13.85 & 17.14 & 26.48 & 19.16 & 11.95 & 9.69 & 7.23\\
Clean ASR Model 	& \checkmark & \checkmark & 11.61 & 13.47 & 16.56 & 23.90 & 17.98 & 11.55 & 9.29 & 6.79\\
\hline
Frequency Domain	& -- & -- & 12.28 & 15.16 & 19.66 & 31.20 & 22.01 & 11.95 & 9.69 & 7.23\\
separation Model 	& \checkmark & -- &12.49 & 15.17 & 19.56 & 30.66 & 21.80 & 7.01 & 2.69 & -1.99\\
+			 	& -- & \checkmark & 12.22 & 13.96 & 16.92 & 25.91 & 18.93 &  11.95 & 9.69 & 7.23\\
Mixture ASR Model 	& \checkmark & \checkmark & 12.41 & 14.77 & 18.83 & 28.99 & 20.86 & 7.93 & 3.88 & -0.75\\
\hline \hline
Time Domain	 	& -- & -- & 11.40 & 14.35 & 18.23 & 28.59 & 20.39 & 20.74 & 18.17 & 15.40\\
separation Model	& \checkmark & -- & 11.26 &  	\textbf{12.90} & 15.80 & 23.13 & 17.28 & 20.65 & 17.94 & 15.07\\
+				& -- & \checkmark & 12.88 & 14.23 & 15.86 & 21.30 & 17.13 &  20.74 & 18.17 & 15.40\\
Clean ASR Model	& \checkmark & \checkmark & 12.81 & 13.62  & 15.05  & 19.30 & 15.99 & 20.65 & 18.03 & 15.20\\
\hline
Time Domain		& -- & -- & 12.32 & 13.95 & 16.72 & 23.43 & 18.03 & 20.74 & 18.17 & 15.40\\
separation Model	& \checkmark & -- & 14.45 & 15.48 & 18.41 & 24.76 & 19.55 & 15.05 & 13.29 & 11.16\\
+ 				& -- & \checkmark & 12.84 & 13.87 & 15.84 & 19.53 & 16.41 & 20.74 & 18.17 & 15.40\\
Mixture ASR Model	& \checkmark & \checkmark & 12.96 & 13.30 & \textbf{15.01} & \textbf{18.31} & \textbf{15.54} & 20.52 & 17.85 & 14.92\\
\hline\hline
\end{tabular}
\vspace{-18pt}
\end{table*}
\vspace{-4pt}
\section{Experimental details}
\vspace{-4pt}
This section presents the experimental details. 
We describe the dataset and the ASR and separation models used for experiments.
We also explain the detail of the joint training. 
We implemented our models in PyTorch.
\vspace{-4pt}
\subsection{Dataset}
\vspace{-4pt}
Experimental mixture data were generated by mixing utterances from speech database with background music.
Both speech and music are sampled at 16 kHz.
As the speech database, we used the Corpus of Spontaneous Japanese Academic Presentation Speech (CSJ-APS).
The CSJ-APS has a duration of around 260 hours and consists of live recordings of academic presentations in nine different academic societies. 
The societies range from engineering, humanities, and social and behavioral sciences.
For background music, we used around 30 hours of background music used in Japanese animations.
\par
For the training dataset, we added background music to the speech with randomly sampled source-to-noise ratio (SNR) levels from a normal distribution with a mean of 0 dB and a standard deviation of 5 dB.
For the test dataset, we added background music from animations not used for the training dataset to the speech of official CSJ-APS testset 1 with various SNR levels such as 5 dB, 0 dB and -5 dB.
This test dataset is referred as the CSJ-anime.
\par
We also evaluated on the dataset mixed with music of particular genres from the Real World Computing Music Database (RWC-MDB) such as classical, jazz and popular. 
Among them, the popular music has lyrics. 
Music signal was added to the CSJ-APS testset speech with SNR levels of 5 dB, 0 dB and -5 dB.
This test datasets are referred as the CSJ-genre.

\vspace{-4pt}
\subsection{Baseline model}
\vspace{-4pt}
We trained two types of baseline ASR models without the separation front-end, which differ in training data; One referred to as clean ASR is trained on clean speech data. The other referred to as mixture ASR is trained on speech-music mixture data.

\vspace{-4pt}
\subsection{Speech-music separation front-end}
\vspace{-4pt}
In this experiment, we compare the following two different speech-music separation networks, which operate in the time-domain and in the frequency-domain, respectively.

\subsubsection{Conv-TasNet network}
We investigated the performance of Conv-TasNet for speech-music separation that uses a similar configuration to the original Conv-TasNet \cite{ConvTasNet}. In particular, following the hyper-parameter notations in the original paper \cite{ConvTasNet}, we set the hyper-parameters N=256, L=20, B=256, H=512, P=3, X=8, R=4.
We used the Adam optimizer to train the network with a learning rate of 1e-3.
This network is referred as the time-domain separation model. 

\subsubsection{Frequency-domain BiLSTM network}
We compare the time-domain separation model with a frequency-domain BiLSTM network, which uses a mask estimation network consisting of five BiLSTM layers with 320 units followed by a linear layer with sigmoid activation.
The input of the mask estimation network consists of an amplitude spectrogram computed with the STFT with a hanning window of 25 ms and a shift of 10 ms.
We reconstructed the waveform of the predicted speech signal by applying the ISTFT to the masked spectrogram of the input mixture. We reused the phase information of the mixture.
We set a learning rate of Adam to 1e-3.
This network is referred as the frequency-domain separation model.

\vspace{-4pt}
\subsection{ASR back-end configuration}
\vspace{-4pt}
The attention-based ASR \cite{AttentionASR} back-end uses two CNN layers with a stride of 2 followed by five BiLSTM layers with 320 units for an encoder, layer normalization \cite{LayerNorm} and two LSTM layers with 320 units for a decoder. 
40-channel lmfb features are used as acoustic features for ASR.
The output of the decoder is a sequence of subwords defined by the byte-pair encoding (BPE) \cite{BPE}. The number of the BPE units is 9,515.

\vspace{-4pt}
\subsection{Joint training}
\vspace{-4pt}
As explained in Section \MakeUppercase{\romannumeral 3}, we design the entire end-to-end model in order that the gradients of the top-level ASR loss can be propagated down to the front-end.
We convert the waveform output of the front-end to a sequence of 40-channel lmfb features before feeding it to the back-end.
We used the STFT implemented in PyTorch through which we can propagate gradients. 
The STFT is set with a hanning window of 25ms and a shift of 10ms which is consistent with feature extraction for pre-training the ASR model.
\par
We compare three different variants of joint fine-tuning: fine-tuning the front-end by propagating gradients through the ASR back-end but only updating the front-end parameters, fine-tuning only the ASR back-end on the enhanced signals, and jointly fine-tuning both components. 
For all variants of joint fine-tuning, the weight for the ASR loss $\alpha$ is set to 2 and a learning rate of Adam is set to 1e-4.
\vspace{-5pt}
\section{Results and Discussion}
\vspace{-4pt}
\subsection{Result of joint fine-tuning}
\vspace{-4pt}
We evaluated the joint model with combinations of two kinds of front-ends of time-domain separation and frequency-domain separation model, and two kinds of back-ends of clean ASR and mixture ASR.
Moreover, fine-tuning was conducted in three cases: the separation front-end only, the ASR back-end only, and both.
We evaluate the performance in terms of source-to-distortion ratio (SDR) \cite{SDR} and word error rate (WER).
The results of these variants are listed for comparison in Table \ref{tb1}.
The comparison among the models is primarily based on the average WER over all SNR levels.
\par
Without any front-end processing, the average WER of clean ASR was 72.97\% and that of mixture ASR was 22.15\%.
It is notable that cascading the independently trained front-end and back-end models already give a better performance than the baseline models whether in the time-domain or in the frequency-domain. 
In this case, the back-end models based on mixture ASR achieve a relative WER reduction of at least 11\% in both domains.
\par
Fine-tuning the ASR back-end while freezing the front-end can further reduce the WER for all combinations of cascading models.
Joint fine-tuning of the frequency-domain separation with clean ASR model achieves the average WER of 17.98\%, which is the best among the frequency-domain methods. 
Joint fine-tuning of the frequency-domain separation with mixture ASR model did not improve from fine-tuning only the ASR back-end and significantly degrades the speech-music separation performance.
In this case, fine-tuning of the front-end makes it unable to separate speech and music because the mixture ASR back-end can already deal with music-mixed speech input.
On the other hand, joint fine-tuning of the time-domain separation with clean ASR and mixture ASR models were both effective, and achieved the average WER of 15.99\% and 15.54\% respectively.
The time-domain separation performance was not degraded so much by combination with mixture ASR in terms of SDR.
The joint fine-tuning with the mixture ASR back-end achieved a relative WER reduction of 29.8\% from the baseline mixture ASR model alone.
This joint model resulted in the best performance (average WER of 15.54\%) among all settings.
\par
Fine-tuning only the separation front-end is not so effective as fine-tuning the ASR back-end for all combinations of the joint models.
It may be because ASR can adapt to the artifacts produced by the front-end.

\subsection{Result of particular music genres}
\vspace{-4pt}
Table \ref{tb2} shows the results of the baseline mixture ASR and the time-domain joint model for unseen genres in the training data.
For all cases of the music genres and SNR levels, our best model improved the WER from the baseline. 
The results on the classical and jazz music datasets show the similar performance to that of the CSJ-anime.
The popular music has lyrics which degrade the performance, but this result still shows that the proposed joint approach is also effective on the data with vocal music.
In general, the results in Table II demonstrate the generalization capability of the proposed method for unseen interferences.

\begin{table}[t]
\small
\centering
\caption{WER and SDR on CSJ-genre for Mixture ASR model and Joint model of Time-domain separation with Mixture ASR}

\label{tb2}
\setlength\tabcolsep{4pt} 
\begin{tabular}{c|c|ccc|ccc}
\multicolumn{8}{r}{WER: Word error rate, SDR: Signal-to-distortion ratio}\\
\hline \hline
&  & \multicolumn{3}{c|}{WER(\%)} & \multicolumn{3}{c}{SDR(dB)} \\
Model & Genre &  5 dB & 0 dB & -5 dB  & 5 dB & 0 dB & -5 dB\\
\hline \hline
		& Classical	& 14.43 & 18.20 & 26.46 &  & & \\
Mixture	& Jazz 		& 14.46 & 17.50 & 25.73 & & & \\
		& Popular 		& 16.12 & 22.64 & 37.18 & &&\\
\hline
Joint		& Classical	& 13.33 & 14.37 & 17.32 & 21.00 & 18.29 & 15.27\\
model	& Jazz 		& 13.29 & 14.43 & 17.48 & 20.83 & 18.05 & 14.94\\
(our best)	& Popular 		& 13.98 & 16.19 & 22.60 & 19.67 & 16.83 & 13.56\\

\hline\hline

\end{tabular}
\vspace{-18pt}
\end{table}

\vspace{-4pt}
\section{Conclusions}
\vspace{-4pt}
We have proposed to combine a time-domain speech-music separation model Conv-TasNet with an attention-based ASR model to form an end-to-end music-mixed speech recognizer.
We show that time-domain separation is better than frequency-domain separation, and pre-training the ASR model on the mixture data is effective.
Joint fine-tuning further significantly improved the performance.
The effectiveness was confirmed with a variety of music genres.

\vspace{-4pt}
\section{Acknowledgements}
\vspace{-4pt}
This research was supported by NII CRIS collaborative research program operated by NII CRIS and LINE Corporation.


\vspace{10pt}
\begin{thebibliography}{1}
\vspace{-4pt}
\bibitem{Vanroose2003}
P. Vanroose and K. Arenberg, ``Blind source separation of speech and background music for improved speech recognition,'' in
\emph{The 24th Symposium on Information Theory,}
2003, pp. 103-108.

\bibitem{Hughes2012}
T. Hughes and T. Kristjansson, ``Music models for music-speech separation,'' in
\emph{2012 IEEE International Conference on Acoustics, Speech and Signal Processing (ICASSP),} Kyoto, 2012, pp. 4917-4920

\bibitem{RobustNMF}
C. Demir, M. Sara\c{c}lar, and A. T. Cemgil, ``Single-Channel Speech-Music Separation for Robust ASR with Mixture Models,'' in 
\emph{IEEE Transactions on Audio, Speech, and Language Processing,} vol. 21, no. 4, pp. 725-736, April 2013

\bibitem{RobustPCA}
P.-S. Huang, S. D. Chen, P. Smaragdis, and M. Hasegawa-Johnson,
``Singing-voice separation from monaural recordings using robust principal component analysis,'' in 
\emph{2012 IEEE International Conference on Acoustics, Speech and Signal Processing (ICASSP),} Kyoto, 2012, pp.
57-60

\bibitem{CDAE}
M. Zhao, D. Wang, Z. Zhang, and X. Zhang, ``Music Removal by Convolutional Denoising Autoencoder in Speech Recognition,'' in 
\emph{2015 Asia-Pacific Signal and Information Processing Association Annual Summit and Conference (APSIPA),} Hong Kong, 2015, pp. 338-341

\bibitem{DAE}
J. Malek, J. Zdansky, and P. Cerva, ``Robust Automatic Recognition of Speech with background music,'' in 
\emph{2017 IEEE International Conference on Acoustics, Speech and Signal Processing (ICASSP),} New Orleans, LA, 2017, pp. 5210-5214

\bibitem{ConvTasNet}
Y. Luo and N. Mesgarani, ``Conv-TasNet: Surpassing Ideal Time-Frequency Magnitude Masking for Speech Separation,'' in 
\emph{IEEE/ACM Transactions on Audio, Speech, and Language Processing,} vol. 27, no. 8, pp. 1256-1266, Aug. 2019



\bibitem{NoiseReduction}
K. Kinoshita, T. Ochiai, M. Delcroix, and T. Nakatani, ``Improving Noise Robust Automatic Speech Recognition with Single-Channel Time-Domain Enhancement Network,'' in 
\emph{2020 IEEE International Conference on Acoustics, Speech and Signal Processing (ICASSP),} Barcelona, Spain, 2020, pp. 7009-7013

\bibitem{SingingVoice}
A. D\'efossez, N. Usunier, L. Bottou, and F. Bach, ``Music Source Separation in the Waveform Domain,''
\emph{arXiv preprint arXiv:1911.13254,} 2019

\bibitem{AttentionASR}
J. K. Chorowski, D. Bahdanau, D. Serdyuk, K. Cho, and Y. Bengio, ``Attention-based models for speech recognition,'' in \emph{Advances in neural information processing systems,} 2015, pp. 577-585.

\bibitem{JointConvTasNet}
T. von Neumann et al., ``End-to-End Training of Time Domain Audio Separation and Recognition,'' in 
\emph{2020 IEEE International Conference on Acoustics, Speech and Signal Processing (ICASSP),} Barcelona, Spain, 2020, pp. 7004-7008

\bibitem{SI-SDR}
J. L. Roux, S. Wisdom, H. Erdogan, and J. R. Hershey, ``SDR --  Half-baked or Well Done?,'' in 
\emph{2019 IEEE International Conference on Acoustics, Speech and Signal Processing (ICASSP),} Brighton, United Kingdom, 2019, pp. 626-630

\bibitem{uPIT}
M. Kolb\ae k, D. Yu, Z. Tan, and J. Jensen, ``Multitalker Speech Separation With Utterance-Level Permutation Invariant Training of Deep Recurrent Neural Networks,'' in 
\emph{IEEE/ACM Transactions on Audio, Speech, and Language Processing,} vol. 25, no. 10, pp. 1901-1913, Oct. 2017

\bibitem{LayerNorm}
J. L. Ba, J. R. Kiros, and G. E. Hinton. ``Layer Normalization,''
\emph{arXiv preprint arXiv:1607.06450,} 2016

\bibitem{BPE}
R. Sennrich, B. Haddow, and A. Birch, ``Neural Machine Translation of Rare Words with Subword Units,'' in 
\emph{Proceedings of the 54th Annual Meeting of the Association for Computational Linguistics (Volume 1: Long Papers),} Berlin, Germany, vol 1, pp.1715-1725, Aug, 2016

\bibitem{SDR}
E. Vincent, R. Gribonval, and C. Fevotte, ``Performance measurement in blind audio source separation,'' in 
\emph{IEEE Transactions on Audio, Speech, and Language Processing,} vol. 14, no. 4, pp. 1462-1469, July 2006



\end{thebibliography}
\end{document}